# Incrementally Maintaining Classification using an RDBMS

M. Levent Koc
University of Wisconsin-Madison
koc@cs.wisc.edu

Christopher Ré
University of Wisconsin-Madison
chrisre@cs.wisc.edu

## ABSTRACT

The proliferation of imprecise data has motivated both researchers and the database industry to push statistical techniques into relational database management systems (RDBMSes). We study strategies to maintain model-based views for a popular statistical technique, *classification*, inside an RDBMS in the presence of updates (to the set of training examples). We make three technical contributions: (1) A strategy that incrementally maintains classification inside an RDBMS. (2) An analysis of the above algorithm that shows that our algorithm is optimal among all deterministic algorithms (and asymptotically within a factor of 2 of a non-deterministic optimal strategy). (3) A novel hybrid-architecture based on the technical ideas that underlie the above algorithm which allows us to store only a fraction of the entities in memory. We apply our techniques to text processing, and we demonstrate that our algorithms provide an order of magnitude improvement over non-incremental approaches to classification on a variety of data sets, such as the Citeseer and DBLife.

## 1. INTRODUCTION

Motivated by the proliferation of imprecise data, there is a growing trend among database researchers and the database industry to push statistical techniques into relational database management systems (RDBMSes). Building on these efforts, our HAZY project aims to build an end-to-end system for imprecision management; a critical portion of our effort is to tightly integrate statistical techniques into an RDBMS. In this work, we focus on the technical challenges that arise when maintaining a ubiquitous statistical task, *classification*, inside an RDBMS.

Classification can be defined as the following task: Given a set of entities $E$ (e.g., papers extracted from the Web), a set of labels $L$ (e.g., research areas), and a set of labeled entities, called training examples, $T$ (e.g., papers labeled by research area), the goal of classification is to assign to each entity in $E$ a label from $L$. Typically, this is done by learning (training) a *model* (which depends on $T$) and then using that model to label each entity in $E$. Classification is widely used, e.g., in extracting structure from Web text [12, 27], in data integration [13], and in business intelligence [7, 14, 22].

Many of these application scenarios are highly dynamic: new data and updates to the data are constantly arriving. For example, a Web portal that publishes information for the research community, it must keep up with the new papers that are constantly published, new conferences that are constantly announced, etc. Similar problems are faced by services such as Twitter or Facebook that have large amounts of user generated content. Unfortunately, current approaches to integrating classifiers with an RDBMS treat classifiers as a data mining tool [22]. In data mining scenarios, the goal is to build a classification model for an analyst, and so classification is used in a batch-oriented manner; in contrast, in the above scenarios, the classification task is integrated into the run-time operation of the application. As many of these applications are often built on RDBMSes, it has now become critical to have classifiers that are integrated with an RDBMS and are able to keep up with high rates of arrival of new information in these dynamic environments.

In the above dynamic applications, there are two types of dynamic data: (1) new entities, e.g., new papers to classify, and (2) new training examples, e.g., as the result of user feedback or crowdsourcing. To handle data of type (1), when a new entity arrives, we must classify the entity according to the model and then store the entity and its label. Handling data of type (2) is more challenging. The reason is that when a new training example arrives, the model itself changes. In turn, this may force us to change the label of every entity in the database. HAZY offers solutions for both problems, but our primary technical focus is on type (2) dynamic data.

We integrate classification into an RDBMS (in our case PostgreSQL) via the abstraction of *model-based views*. In model-based views, statistical computations are exposed to a developer through relational views. As we explain in Section 2.1, this abstraction allows a developer to use the output of classification with standard SQL queries, and allows new training examples to be inserted with standard SQL insert statements. Thus, to a developer that uses HAZY, the result of classification appears to be a standard relational view.

In full analogy with classical materialized views [3], Deshpande and Madden [11] considered two approaches to maintaining general statistical views: *eager*, where the view is maintained after each update (i.e., new training example), and *lazy*, where updates are applied only in response to a read request (i.e., the request for the label of an entity). The eager approach allows very high read performance, but has higher maintenance costs; in contrast, the lazy approach offers more efficient updates, but has worse read performance. In this work, we describe a novel incremental data reorganization strategy that can improve the performance of either an eager or lazy approach.

To gain intuition, we explain how HAZY improves the update performance of an eager approach: on each update, a naïve eager approach relabels every entity. Intuitively, this operation is wasteful: although the labels of many entities do not change as a result of a single new training example, an eager approach must still read every entity tuple, and write back each entity that changed label. If, however, we could quickly identify the set of entities whose labels change (or may change), then we could save work by reading and writing a smaller number of entities on each update. To find the entities that are likely to change labels, HAZY provides a data structure and a strategy that clusters together entities on disk by how likely their label is to change. Using this clustering, HAZY can process dramatically fewer tuples and so can achieve much higher performance.

Maintaining this clustering as new updates arrive raises an immediate technical problem. Intuitively, the reason is that the set of entities whose labels are likely to change may not remain the same after many updates. To continue to reap the benefits of the clustering, we need to periodically reorganize the data. Our first contribution is a strategy that chooses when to reorganize the data. At a high level, our strategy weighs the expected benefit of reorganization against the time that it takes to reorganize the data set: if reorganization is cheaper in this calculation, then it chooses to reorganize. Our second contribution is our analysis that shows that HAZY's maintenance strategy has an optimal run time among all deterministic strategies for this problem and is (asymptotically in the size of the data) a 2-approximation of a nondeterministic optimal strategy (see Thm. 3.3). Empirically, HAZY's strategy results in an order of magnitude performance improvement over naïve approaches.

Using this strategy, we design both main-memory and on-disk architectures for view maintenance and integrate them into PostgreSQL. Of course, the main-memory approach is orders of magnitude faster than a naïve on-disk approach, but has the limitation that the data set must fit in main memory. In some of our target applications, this main-memory limitation is a problem, e.g., in text processing a common approach is to have a vector for each document with a component for each word in the corpus: even represented using sparse vectors, millions of entities may still require tens of gigabytes of RAM. To combat this, our third technical contribution is a hybrid of main-memory and on-disk architectures that keeps some number of entities in main memory: the technical problem is: given a space budget, which entities do we keep in main memory? At a high level, HAZY's strategy tells us which fraction of the entities are likely to change labels, and the hybrid architecture stores these entities in memory. Often, we can store a small fraction of the entities in memory (say 1%) and still service almost every read and update without going to disk. Combining both our architectural techniques and algorithmic techniques, we show up to two orders of magnitude performance improvements over state-of-the-art techniques.

### Prior Work on Statistical Views in RDBMSes

Pushing statistical processing (e.g., SVMs [22]) inside an RDBMS is a trend in the database industry. No prior approaches, however, consider efficient strategies for maintenance of classification views. Our work in HAZY extends the line of work started by MAUVEDB [11] that proposes model-based views and considers materialization strategy for such views, i.e., should the data be materialized lazily, partially, or eagerly? This concept has been expanded to statistical views for probabilistic inference over *Markovian Sequences* [17,23] where all updates are appends. The Monte-Carlo Database System [15] makes statistical sampling a first class primitive, but Jampani *et al.* do not discuss maintaining statistical views. A related line of work in *probabilistic databases* [1,2,10] considers specialized models to manage uncertainty, in this work we address the orthogonal problem of building a statistical model from raw data.

HAZY builds on work in the machine learning community to incrementally learn support vector machines and linear models [6,9,19,21]; this work is essentially used as a subroutine in HAZY. In HAZY, our goal is to address the problem of incrementally maintaining the output of a classification task in an RDBMS. For an expanded related work, please see Appendix D.

### Contributions, Validation, and Outline

- In Section 2, we illustrate HAZY's use by example, define its semantics, and give the necessary background on lazy and eager approaches.
- In Section 3, we describe our main technical contributions: an efficient strategy to support incremental maintenance for classification views, a theoretical analysis of our strategy, and several architectural optimizations based on our strategy.
- In Section 4, we validate our technical contributions on several real data sets from DBLife, Citeseer, and the UCI machine learning repository.[1] We demonstrate that HAZY's incremental strategy is an order of magnitude more efficient than non-incremental strategies on several data sets.

In the appendix, we include pseudo code, proofs of our claims, an extended related work, and extended experiments including how we use HAZY in multiclass classification.

## 2. EXAMPLE AND BACKGROUND

We introduce HAZY by example and conclude this section with a formal description of HAZY's semantics.

### 2.1 Hazy by Example

We describe how a user declares a classification view in HAZY by example.

*Feature Functions.* Before defining HAZY's views, we need the concept of a *feature function*, which is a user-defined function that maps an (entity) tuple to a vector describing the features that are important for a classification task. An example feature function is `tf_bag_of_words` that computes a vector of the term (word) frequencies for each tuple (treating each tuple as a document). We discuss HAZY's registration of feature functions in Appendix A.2.

---
[1] http://archive.ics.uci.edu/ml/

*Classification Views.* To create a list of newly published database systems papers for a Web page, we need to separate papers about database systems from those papers about other research areas. In HAZY, we start by declaring a *classification view*, which contains a declarative specification of the entities and the training examples. HAZY populates and maintains the view by performing the classification task.

**Example 2.1** Classifying papers by area in HAZY:

```
CREATE CLASSIFICATION VIEW
 Labeled_Papers KEY id -- (id,class)
 ENTITIES FROM Papers KEY id  -- (id, title, ...)
 LABELS   FROM Paper_Area LABEL l -- (label)
 EXAMPLES FROM Example_Papers KEY id LABEL l -- (id, label)
 FEATURE FUNCTION tf_bag_of_words
```

Here, we have illustrated the schema of each portion of the view with SQL comments. Example 2.1 declares a view, `Labeled_Papers`, with two attributes (id, class) that contains each paper in `Papers`, but now labeled by the classification mechanism, e.g., a paper may be labeled as a database paper. The developer may access `Labeled_Papers` as a standard SQL table (although HAZY is responsible for maintaining this view). This also declares to HAZY that there is an existing table (or view) `Papers` that contains the entities (papers) to be labeled and has a primary key id. `Paper_Area` declares a set of labels (here the set of research areas). Finally, the training examples are specified by a table `Example_Papers` with schema (id, label) where id refers to a paper, and label is one of the labels previously declared. A developer may insert new training examples into this table, and HAZY will ensure that the `Labeled_Papers` view reflects the result of retraining the model with these examples. Of course, one option is to simply redo training and then populate the view. An alternative is to incrementally maintain the view by leveraging incremental algorithms [6, 9, 19]; these algorithms tell us how to update the classifier's model, but they do not tell us how to efficiently reclassify the entities inside the database. Our first technical contribution is an *incremental* algorithm to maintain classification views as the underlying training example (and so model) is updated.

In HAZY, classification may be implemented in several ways: *support vector machines* [5, 8, 16], *ridge regression, ordinary and weighted least squares*, or *logistic regression*; each of these methods can also be combined with various *kernels* (see Appendix B.5.2). In HAZY a user specifies one of a handful of predefined classification methods in the view declaration, e.g., by writing `USING SVM` at the end of the view declaration. If the user does not specify, HAZY chooses a method automatically (using a simple model selection algorithm based on *leave-one-out-estimators* [26, p. 27]).

*Using* HAZY. A key virtue of HAZY's approach is that a user may write arbitrary SQL in terms of a classification view. This is trivial to support, because classification views appear to the RDBMS as exactly standard database views. To support user feedback, HAZY allows standard SQL updates, inserts, and deletes to the tables (or views) that define the entities and training examples of a classification view. Using triggers, HAZY monitors the relevant views for updates, and the challenge for HAZY is to incrementally maintain the contents of a classification view.

*Semantics of the Views.* We define the contents of a classification view. To keep our formal presentation concise, we focus on a classification view that is defined by a *linear classifier*, which is a very broad class containing linear support vector machines, least squares, and logistic regression. HAZY does handle non-linear kernels and multiclass problems (for details see Appendix B.4). Formally, a linear model is defined by a pair $(\boldsymbol{w}, b)$ called a *model* where $\boldsymbol{w} \in \mathbb{R}^d$ for $d \geq 1$ and $b \in \mathbb{R}$ ($b$ is called the *bias term*).

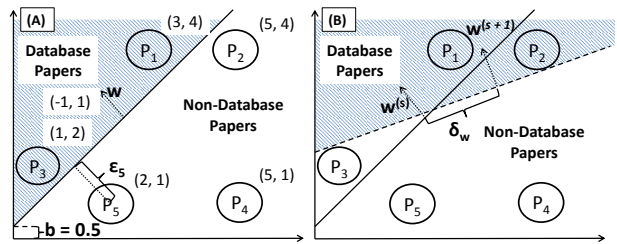

**Figure 1: An Example Linear Model**

A classification view $V(\text{id}, \text{class})$ is defined by a pair $(\text{In}, \text{T})$ where $\text{In}(\text{id}, \text{f})$ is a view with a key *id* that defines the set of entities to be classified and their feature vectors f. Here $\text{f} \in \mathbb{R}^d$ for $d \geq 1$, and contains the result of applying the feature function to the tuple with key id. $T(\text{id}, \text{l})$ is a view that defines the training examples: id is a key, and $\text{l} \in \{-1, +1\}$ is the label. A model $(\boldsymbol{w}, b)$ is a solution to an optimization problem that depends only on T (see Appendix A.1). In turn, a model $(\boldsymbol{w}, b)$ defines the contents of $V$ as:

$$V = \{(id, c) \mid (id, f) \in \text{In and } c = \text{sign}(\boldsymbol{w} \cdot f - b)\}$$

where $\text{sign}(x) = 1$ if $x \geq 0$ and $-1$ otherwise.

**Example 2.2** A simple linear model is illustrated in Figure 1 that labels entities as database papers or not. There are five entities here, $P_i$ for $i = 1, \ldots, 5$ that are mapped by feature functions to vectors in $\mathbb{R}^2$. For example, $P_1$ is associated with the vector $f(P_1) = (3, 4)$. The model in the example is the pair $(\boldsymbol{w}, b)$ where $\boldsymbol{w} = (-1, 1)$ and $b = 0.5$. The model vector $\boldsymbol{w}$ defines a line by the set of points $\boldsymbol{x} \in \mathbb{R}^2$ such that $\boldsymbol{x} \cdot \boldsymbol{w} + b = 0$: those entities on above the line are in the class (database papers) and are shaded, and those below the line are not. In higher dimensions, the line is generalized to a hyperplane. The shaded region is the set of $\boldsymbol{y}$ such that $\boldsymbol{y} \cdot \boldsymbol{w} + b \geq 0$, and we call this set the positive class. Hence, according to this model, both $P_1$ and $P_3$ are labeled as database papers, while $P_j$ for $j \in \{2, 4, 5\}$ are not.

In HAZY, classification views are defined using standard SQL views, e.g., in Example 2.1, In is defined by the SQL view `Papers`, while the training examples are defined by the view `Example_Papers`. HAZY is responsible for maintaining $V$; this paper studies the mechanisms to maintain $V$ efficiently as the model changes. As updates arrive (inserts or deletes) the model changes – but perhaps only slightly as the result of incremental training: Figure 1(b) shows a model $w^{(s)}$ (at time $s$) and then it changes to a new model $w^{(s+1)}$ using an incremental training algorithm. Only $P_2$ and $P_3$ change classes here, and it is up to HAZY to quickly identify $P_2$ and $P_3$ and relabel only such entities.

## 2.2 Background: Lazy versus Eager

HAZY handles arbitrary SQL, but for the sake of presentation, we focus on three operations: (1) *Single Entity* read,

which is a query that asks for the label of a single entity, e.g., *"is paper 10 a database paper?"*, (2) *All Members*, which is a query that asks for all members of a class, e.g., *"return all database papers"*, and (3) *Update*, which adds a new training example, e.g., *"paper 45 is a database paper."*[2]

We describe the eager and lazy approaches in more detail. The input is a table In(id, f), where id is a key and f is a feature vector (it is an element of $\mathbb{R}^d$ for some $d \geq 1$). The goal is to maintain a view $V(\text{id}, \text{class})$. In an eager approach, $V$ is maintained as a materialized table. In contrast, in a lazy approach, $V$ is implemented using an SQL view whose definition contains a user-defined function that in response to a read of a tuple with id $= x$, reads the feature vector associated to $x$ in In, and labels it using the current model. In both cases, we maintain a hash index to efficiently locate the tuple corresponding to the single entity.

*Naïve Approach.* The naïve approach for *Single Entity* read in an eager approach is to simply look up the result in the materialized view $V$. In a lazy approach for a *Single Entity* read, we must retrieve the tuple and its feature vector, and then classify the tuple. For an *All Members* read, in an eager approach we simply scan the table looking for those tuples in the class, while in a lazy approach we must scan and relabel every single tuple. On an *Update*, both approaches first retrain the model. On our data sets, retraining the model takes roughly on the order of 100µs. In a lazy approach, this is all the work that the update needs to do. In contrast, in an eager approach we must reclassify all the entities which requires an entire scan over the corpus and update the documents. A technical intuition behind HAZY is that much of the work in this scan is wasted – often only a small fraction of labels change on any given iteration.

*Hazy's Technical Goals.* HAZY's techniques can improve either eager or lazy approaches. Specifically, HAZY improves the cost of *Update* in the eager approach, while maintaining its *All Members* performance. (HAZY slightly improves *All Members* as well); in the lazy approach, HAZY improves the *All Members* query while retaining the lazy approach's optimal *Update* performance. We also propose a hybrid architecture that improves *Single Entity* performance.

## 3. CLASSIFICATION VIEWS IN HAZY

We describe how HAZY initially trains a model, HAZY's novel strategy for incrementally maintaining classification views, and several optimizations.

### 3.1 Initialization (Learning)

HAZY is agnostic about the particular choice of learning algorithm, but HAZY's default learning algorithm is *stochastic gradient* [4], an incremental optimization technique. We choose this as HAZY's default, since incremental gradient algorithms examine a small number of training examples per step, and so they have a small memory footprint during the training phase [9, 25]. Additionally, these methods allow HAZY to efficiently and incrementally maintain the model at each step. Although this method is simple, as we demonstrate (Appendix C.1), our approach is substantially faster than other state-of-the-art approaches (namely, SVMLIGHT and an established commercial system), while maintaining as good (if not better) quality. Moreover, using a gradient-based approach allows HAZY to handle many linear models using essentially the same code.

---
[2]HAZY supports deletion and change of labels by retraining the model from scratch, i.e., not incrementally.

### 3.2 Incremental Maintenance

A challenging operation is to maintain the view when new training examples arrive: once the classification view is constructed, updates to the training examples may cause the model to change and so the contents of the view. In this section, we first explain how HAZY can be applied in the eager setting to speed up *Update* performance. In Section 3.4, we discuss how HAZY improves the performance of *All Members* in a lazy approach.

**Input and Output** We are given as input a relation $\text{In}(id, \text{f})$ where f is a feature vector. Then, in an online fashion we are given a sequence of models $(\boldsymbol{w}^{(i)}, b^{(i)})$ for $i = 1, 2, \ldots$ (we call $i$ a round). The sequence of models capture the result of incrementally training a model with a sequence of examples. At each round $i$, HAZY maintains a (materialized) view $V^{(i)}(\text{id}, \text{class}, \text{eps})$. We abuse notation by redefining $V^{(i)}$ to have an additional attribute eps. HAZY will use eps to ensure sequential access, but it is not needed (and not present) in naïve strategies. The content of $V^{(i)}$ is the set of $(id, c, eps)$ where:

$$(id, f) \in \text{In}, \quad eps = \boldsymbol{w}^{(i)} \cdot f - b^{(i)} \text{ and } c = \text{sign}(eps)$$

Internally, HAZY maintains a scratch table H: for $s \leq i$ the contents of $\text{H}^{(s)}(\text{id}, \text{f}, \text{eps})$ are

$$\text{H}^{(s)} = \{(id, f, eps) \mid (id, f) \in \text{In}, \text{ and } eps = \boldsymbol{w}^{(s)} \cdot f - b^{(s)}\}$$

HAZY maintains that H is clustered on eps. $\varepsilon$ for an entity is the distance to the hyperplane defined by $(\boldsymbol{w}^{(s)}, b^{(s)})$. Figure 1(A), illustrates $\varepsilon$ for paper 5.

**Algorithm Overview** At round $i+1$, HAZY has a single instantiation of $\text{H}^{(s)}$ for some $s \leq i$ ($s$ will signify the last round HAZY reorganized the data). The algorithm has two components: The first component is an *incremental step* that uses $\text{H}^{(s)}$ to update a (hopefully small) portion of $V^{(i)}$ to build $V^{(i+1)}$. Intuitively, after many rounds (when $i$ is much bigger than $s$), the model at $s$ and $i$ are far apart, and so the incremental step may not perform well. And so, the second component of HAZY's strategy is deciding when to reorganize H. We call the (online) algorithm that chooses when HAZY reorganizes the data the SKIING strategy (referring to the classic *ski rental problem* [18]).

#### 3.2.1 The Skiing Strategy

The SKIING strategy operates in rounds. In each round, say $i + 1$, SKIING makes one of two choices: (1) perform an incremental step and pay a cost $c^{(i)}$ to update $V^{(i)}$ to $V^{(i+1)}$. The cost $c^{(i)}$ is in seconds ($c^{(i)}$ is unknown to the strategy until it makes its choice). Or (2) a reorganization step that reorganizes H (effectively setting $s = i + 1$) and pays a fixed, known cost $S$ (and also updates $V^{(i)}$). We describe this formally in Appendix 3.3.

Let $\alpha \geq 0$ be a constant ($\alpha = 1$ suffices in practice). When SKIING makes choice (1) it measures the time to execute the step in round $i$ and calls this cost $c^{(i)}$, then SKIING maintains an accumulated cost $a^{(i)}$ (with an optimal data organization SKIING would incur cost 0, and so intuitively, $a^{(i+1)}$ represents the waste incurred by SKIING's most recent

305

| | id | eps | f | |
|---|---|---|---|---|
| $\geq hw^{(s,i)}$ | 10 | 0.4 | $(3:0.1,\ldots$ | $V_+^{(i)} \supseteq \{10,1\}$ |
| | 1 | 0.2 | $(5:0.2,\ldots$ | $V_+^{(i)} \subseteq \{10,1,34,5,3\}$ |
| To be reclassified | 34 | 0.1 | $(1:0.5,\ldots$ | |
| | 5 | 0.01 | $(3:0.9\ldots$ | |
| | 3 | -0.2 | $(185:0\ldots$ | $V_-^{(i)} \supseteq \{18\}$ |
| $\leq lw^{(s,i)}$ | 18 | -0.5 | $(6:0\ldots$ | $V_-^{(i)} \subseteq \{34,5,3,18\}$ |
| Low and high water : $lw^{(s,i)} = -0.22$ and $hw^{(s,i)} = 0.15$ | | | | |

**Figure 2: An illustration of $\mathrm{H}^{(s)}$ at a round $i$ where $s \leq i$. Only those tuples (shown in gray) in $[lw^{(s,i)}, hw^{(s,i)}]$ need to be reclassified.**

choice of data organization) as:

$$a^{(i+1)} = a^{(i)} + c^{(i)} \text{ s.t. } a^{(0)} = 0 \qquad (1)$$

When $a^{(i)} \geq \alpha S$, HAZY reorganizes the data (choice (2)) and resets the accumulated cost to 0. The strategy is summarized in Figure 7 in Appendix B.3.

### 3.2.2 The Individual Steps

We now describe the state that HAZY maintains, the incremental step, and the reorganization step.

*State Maintained by Skiing.* At round $i$, HAZY maintains a model $(\boldsymbol{w}^{(s)}, b^{(s)})$ for some $s \leq i$ called the *stored model*. The SKIING strategy chooses when HAZY changes the stored model by choosing the parameter $s$. For convenience, denote the positively labeled elements at round $i$ as $V_+^{(i)} = \{id \mid (id, 1) \in V^{(i)}\}$, and define $V_-^{(i)}$ analogously. HAZY maintains a pair of scalars that depend on the model at round $i$ and the last round HAZY reorganized, $s$; these scalars are denoted $lw^{(s,i)}$ and $hw^{(s,i)}$ (defined below). The important property is that if $t \in \mathrm{H}^{(s)}$ and $t.\text{eps} \geq hw^{(s,i)}$ (resp. $t.\text{eps} \leq lw^{(s,i)}$) then $t.\text{id} \in V_+^{(i)}$ (resp. $t.\text{id} \in V_-^{(i)}$). We informally refer to $lw^{(s,i)}$ and $hw^{(s,i)}$ as the low water and high water, respectively. These scalars are used to form a sufficient condition to determine class membership. This test is incomplete: for those tuples, with $t.\text{eps} \in [lw^{(s,i)}, hw^{(s,i)}]$, the label of $t.\text{id}$ may be either 1 or $-1$. Notice $t.\text{eps}$ is a function of the *stored model* (so $s$), while $t \in V^{(i)}$ depends on the model at round $i$. Figure 2 shows H and the above scalars during execution.

*Incremental Step.* The input at round $i+1$ is a model, $(\boldsymbol{w}^{(i+1)}, b^{(i+1)})$, and our goal is to update the label of the tuples in the view $V^{(i)}$ to build $V^{(i+1)}$. Intuitively, if the model at round $i$ is close (in some norm) to the model that is used to cluster $\mathrm{H}^{(s)}$, then we can use the clustering of $\mathrm{H}^{(s)}$ to examine fewer tuples to build $V^{(i+1)}$.

Given that the model has changed, we need to bound how much closer each entity could have gotten to the hyperplane associated with the new model (c.f. Figure 1(B)). For each feature vector $\boldsymbol{y}$, the technical key is to bound the size of the dot product $|\boldsymbol{w} \cdot \boldsymbol{y}|$ in terms of the length (norm) of $\boldsymbol{w}$ and $\boldsymbol{y}$. There is a beautiful basic inequality from functional analysis called Hölder's inequality that allows us to do just this. Fix any $p, q \geq 0$ such that $p^{-1} + q^{-1} = 1$, called *Hölder conjugates* [24, p.139]. And let $M = \max_{t \in \text{In}} \|t.f\|_q$.

Let $j \geq s$ be a round, we define scalars $\varepsilon_{high}^{(s,j)}$ and $\varepsilon_{low}^{(s,j)}$ as

$$\varepsilon_{high}^{(s,j)} = M\|\boldsymbol{w}^{(j)} - \boldsymbol{w}^{(s)}\|_p + b^{(j)} - b^{(s)}$$
$$\varepsilon_{low}^{(s,j)} = -M\|\boldsymbol{w}^{(j)} - \boldsymbol{w}^{(s)}\|_p + b^{(j)} - b^{(s)}$$

Notice that $M$ is only a function of the set of entities $E$.

LEMMA 3.1. *For $j \geq s$, let $t \in \mathrm{H}^{(s,j)}$. If $t.\text{eps} \geq \varepsilon_{high}^{(s,j)}$ then $t.\text{id} \in V_+^{(j)}$. If $t.\text{eps} \leq \varepsilon_{low}^{(s,j)}$ then $t \in V_-^{(j)}$.*

We use this lemma to define a pair of scalars $lw^{(s,j)}$ and $hw^{(s,j)}$ (low and high water) for $s \leq j$ as follows:

$$lw^{(s,j)} = \min_{l=s,\ldots,j} \varepsilon_{low}^{(s,l)} \text{ and } hw^{(s,j)} = \max_{l=s,\ldots,j} \varepsilon_{high}^{(s,l)} \qquad (2)$$

The only tuples whose label may differ from their setting at $s$ (the last reorganization) have $t.\text{eps} \in [lw^{(s,j)}, hw^{(s,j)}]$. HAZY saves over an eager approach in that it only needs to reclassify tuples satisfying $t.\text{eps} \in [lw^{(s,j)}, hw^{(s,j)}]$ (which may be much smaller than the entire data set). To quickly identify these tuples on disk, HAZY maintains a clustered $B$+-tree index on $t.\text{eps}$ in H.

*Reorganization Step.* Intuitively, for the incremental step to have good performance in round $i$, the stored model, $(\boldsymbol{w}^{(s)}, b^{(s)})$, and the current model, $(\boldsymbol{w}^{(i)}, b^{(i)})$, must be close. After many rounds, these models may drift away from one another. When SKIING tells HAZY to reorganize, HAZY sets the stored model to $(\boldsymbol{w}^{(i)}, b^{(i)})$, reclusters H, and rebuilds the needed indexes; HAZY sets $S$ to the time it takes to perform these operations.

*Choosing the Norm.* In HAZY, the norm ($p$ above) is chosen for quality – not performance – reasons. In text processing $\ell_1$ normalization is used to compensate for documents of different lengths, and so $(p = \infty, q = 1)$ is used. Some applications use $\ell_2$ normalization, and so $(p = 2, q = 2)$ [20].

### 3.3 Analysis of Skiing

We show that SKIING is a 2-approximation (in terms of cost) of an optimal strategy to maintain HAZY's state (in the size of the data), and SKIING is optimal among all deterministic online strategies (up to sub-constant factors).

We make two assumptions. The first is that the cost we pay at $i$ is only a function of the last round HAZY reorganized, and we define $c^{(s,i)}$ to be the cost we pay at round $i$ if our last reorganization was at $s$ ($s \leq i$). Second, we assume that $c^{(s,i)} \leq c^{(s',i)}$ whenever $s \geq s'$, i.e., reorganizing more recently does not raise the cost. In HAZY, both of these assumptions hold as the cost is a function of the number of tuples in $[lw^{(s,i)}, hw^{(s,i)}]$, which is monotone increasing in $i$ for a fixed $s$ (see Eq. 2). In Appendix B.3, we consider alternative incremental strategies that do not satisfy the second property.

Formally, we consider the following on-line problem: At each round $i$, we either (1) *reorganize* and pay cost $S$, or (2) we pay a cost $c^{(s,i)} \leq S$ where $s$ is the last round that the strategy chose to reorganize. Let $\sigma \geq 0$ such that $\sigma S$ is the time to scan H.

We define the notion of a *schedule* to capture the action of any strategy. Fix $N$, the number of rounds, then a *schedule* is an increasing sequence of integers $\bar{u}$ where $0 = u_0 < u_1 < u_2 < \cdots < u_M \leq N$ (here $u_i$ is the round of the $i^{\text{th}}$ reorganization) and $M$ is the number of reorganizations. Given a schedule $\bar{u}$, let $\lfloor i \rfloor_{\bar{u}} = \max \{u \in \bar{u} \mid u \leq i\}$, i.e., the largest element of $\bar{u}$ that is smaller than $i$; intuitively, $\lfloor i \rfloor_{\bar{u}}$ is the most recent round that we reorganized before $i$ in the schedule $\bar{u}$. Denote the set of costs by $\bar{c} = \{c^{(s,i)}\}$ for $s \leq i \leq N$ (satisfying the above properties). The cost of a schedule $\bar{u}$ with length $M$ and reorganization cost $S$ with costs $\bar{c}$ is

$$\text{Cost}(\bar{u}, S, \bar{c}) = \sum_{i=1,\ldots,N} c^{(\lfloor i \rfloor_{\bar{u}}, i)} + MS$$

This says that we pay for $M$ reorganizations (a cost of $MS$) and then we pay for each of the incremental steps.

A strategy $\Psi$ is a function that takes $\bar{c}$ as input and produces a schedule $\bar{u}$. For example, $\bar{u} = \text{SKIING}(\bar{c})$ denotes that $\bar{u}$ is the sequence of reorganizations that the SKIING strategy performs. This definition allows very powerful strategies (e.g., that can see the future). Intuitively, a strategy is a deterministic online strategy if the schedule depends only on the input it has observed up to any point. Formally, $\Psi$ is a deterministic online strategy if $\bar{u} = \Psi(\bar{c})$ and for any other $\bar{d}$ such that $c^{(\lfloor i \rfloor_{\bar{u}}, i)} = d^{(\lfloor i \rfloor_{\bar{u}}, i)}$ we have that $\bar{u} = \Psi(\bar{d})$. The SKIING strategy is a deterministic online strategy: it does not have access to elements of $\bar{c}$ until after it makes its decision in each round.

We define OPT to be a strategy that finds a best (lowest cost) schedule for any set of costs. Our formal statement concerns the *competitive ratio*, denoted $\rho$, between the cost of what a strategy produces and the optimal cost. Let $C$ be the set of sets of costs. For any strategy $\Psi$ define $\rho$ as:

$$\rho(\Psi) = \sup_{\bar{c} \in C} \frac{\text{Cost}(\bar{u}, S, \bar{c})}{\text{Cost}(\bar{o}, S, \bar{c})} \text{ where } \bar{u} = \Psi(\bar{c}), \bar{o} = \text{OPT}(\bar{c})$$

LEMMA 3.2. *Let $\alpha$ (the SKIING parameter) be the positive root of $x^2 + \sigma x - 1$. With the notation above, $\rho(\text{SKIING}) = (1 + \alpha + \sigma)$. Moreover, for any deterministic online strategy $\Psi$, we have $\rho(\Psi) \geq (1 + \alpha + \sigma)$.*

We include the proof in Appendix B.3. Two immediate consequences of this lemma are (1) the SKIING strategy is optimal among all deterministic online strategies, and (2) since $\sigma \to 0$ as $|\text{In}| \to \infty$ (reorganizations contain a sort, which is asymptotically more expensive than a scan); and observing that this implies that $\alpha \to 1$, we have the following asymptotic statement:

THEOREM 3.3. *As the data size grows, $|\text{In}| \to \infty$, and so $\sigma \to 0$, we have that $\rho(\text{SKIING}) \to 2$ and this is optimal among all deterministic online strategies.*

### 3.4 Lazy Hazy

In a lazy approach, the *All Members* query performance is suboptimal, and HAZY improves its performance. When responding to an *All Members* query, HAZY needs to read those tuples above low water ($lw^{(s,i)}$) and so it reads $N_R$ tuples where $N_R \geq N_+$ ($N_+$ is the number in the class). In contrast, with an optimal data layout we would need to read only $N_+$ and so a fraction of this work is wasted. Thus, if the read takes $S$ seconds then HAZY sets $c^{(i)} = (N_R - N_+)N_R^{-1}S$ and updates the accumulated cost $a^{(i)}$ as before. One can check that the formal conditions of Lemma 3.2 are met, and so the same performance guarantees still hold. In a lazy update we do not accumulate waste, since a lazy update has optimal performance, data reorganization cannot help. The rest of the strategy is identical to the eager case.

### 3.5 Architectural Optimizations

We describe two performance enhancements to the on-disk architecture of HAZY: a main-memory architecture, which provides excellent performance when all entities fit in memory, and a hybrid architecture that given a memory budget improves *Single Entity* read rates.

#### 3.5.1 In-memory Architecture

We observe that the output of classification data is a function of the training examples and the entities. Since we can always recompute the model, we only need to store the training examples and entities persistently. Thus, we can maintain the classification view in memory for performance, and when memory needs to be revoked we do not have to write it back to disk – we can simply discard it. We leverage this observation to make an improved main memory architecture that we refer to as HAZY-MM. The structure we maintain in memory is similar to the structure in Section 3.2, e.g., we still cluster the data in main memory, which is crucial to achieve good performance, and we use the SKIING strategy to choose when to reorganize the data.

#### 3.5.2 Hazy's Hybrid Architecture

The goal of our hybrid solution is to reduce the memory footprint of the main-memory architecture, while improving the *Single Entity* read rate of both the lazy and eager approaches. We again use the SKIING strategy to reorganize the data. We continue to maintain the on-disk structure previously described. In memory, we maintain a buffer of $B$ entities, and a hash map $h^{(s)} : \text{E} \to \mathbb{R}$ that we call the $\varepsilon$-map defined as follows: for each $t \in \text{H}^{(s)}$ we define $h^{(s)}(t.\text{id}) = \varepsilon$ where $\varepsilon = t.\text{eps}$. In addition, we continue to maintain $lw^{(s,i)}$ and $hw^{(s,i)}$. When $h^{(s)}(t) \geq hw^{(s,i)}$ (resp. $h^{(s)}(t) \leq lw^{(s,i)}$), then we are sure that $t.\text{id} \in V_+^{(i)}$ (resp. $t.\text{id} \in V_-^{(i)}$). In this case, we simply return the correct value. Otherwise, we check the buffer, and if it is not present in the buffer, then HAZY goes to disk. Appendix B.4 includes pseudocode for the search procedure. Calculating $c^{(i)}$ for eager and lazy approaches in the hybrid architecture is a straightforward extension of the previous architectures. Again, the SKIING strategy reorganizes the data on disk and in memory.

The main-memory footprint of our hybrid architecture may be much smaller than the total size of the data: it is $B(\mathsf{f} + \mathsf{k}) + (\mathsf{k} + \text{sizeof}(double))N$ where $\mathsf{k}$ is the size of the primary key in bytes, $\mathsf{f}$ is the size of the feature function in bytes, and $N$ is the number of tuples in the data set; in contrast, the whole data set size is $N(\mathsf{k} + \mathsf{f})$. On text data, $\mathsf{f}$ may be in the thousands of bytes, e.g., each word in the abstract may be a feature. For example, the Citeseer data set (with feature vectors) is 1.3GB, yet the $\varepsilon$-map is only 5.4MB (over 245x smaller).

## 4. EXPERIMENTS

We verify that HAZY's techniques provide performance gains across various classification tasks.

*Prototype Details and Experimental Setup.* HAZY is implemented in PostgreSQL 8.4. We use a linear support vector machine for classification. We have implemented several gradient and incremental training algorithms, but we perform all experiments in this section with *stochastic gradient* based on Bottou's code [4] in C. All updates to the classi-fication views are monitored using standard triggers; HAZY runs in a separate process and IPC is handled using sockets. Further details can be found in Appendix B.1. Unless otherwise noted, all experiments are run on a single configuration: Intel Core2 CPU at 2.4Ghz with 4GB of RAM and two SATA disks running 64-bit Linux 2.6.18-194. Performance numbers are run 7 times and the times are averaged. For uniformity, we set $\alpha = 1$ for all experiments (Section 3.2.1).

| Data set | Abbrev | Size | # Entities | $|F|$ | $\neq 0$ |
|---|---|---|---|---|---|
| Forest | FC | 73M | 582k | 54 | 54 |
| DBLife | DB | 25M | 124k | 41k | 7 |
| Citeseer | CS | 1.3G | 721k | 682k | 60 |

**Figure 3: Data Set Statistics.** For each data set, we report its size in Bytes, the number of entity references in the data set, and the number of features.

| Technique | | FC | DB | CS | FC | DB | CS |
|---|---|---|---|---|---|---|---|
| OD | Naive | 0.4 | 2.1 | 0.2 | 1.2 | 12.2 | 0.5 |
|    | Hazy  | 2.0 | 6.8 | 0.2 | 3.5 | 46.9 | 2.0 |
| Hybrid | | 2.0 | 6.6 | 0.2 | 8.0 | 48.8 | 2.1 |
| MM | Naive | 5.3 | 33.1 | 1.8 | 10.4 | 65.7 | 2.4 |
|    | Hazy  | 49.7 | 160.5 | 7.2 | 410.1 | 2.8k | 105.7 |
| | | Eager *Update* (Updates/s) (A) | | | Lazy *All Members* (Scan/s) (B) | | |

**Figure 4: Eager Updates and Lazy All Members.**

*Data set Descriptions.* In the performance experiments we use three data sets: Citeseer (CS), Forest (FC), and DBLife (DB). Forest is a standard benchmark machine learning data set with dense feature vectors.[3] DBLife is the set of papers crawled by the DBLife Web portal. Our goal is to classify these papers as database papers or not. Citeseer is a large set of papers that we also want to classify as database papers or not. In both DBLife and Citeseer, we use a bag of words feature set and so use sparse feature vectors. A key difference between Citeseer and DBLife is that we use the abstracts as features in Citeseer (as opposed to simply the title in DBLife), this results in more non-zero components in the sparse feature vector representation. Statistics for the data sets that we use are shown in Figure 3. We include accuracy results and bulk-loading times in Appendix C.1.

### 4.1 Maintaining Classification Views

The central claim of this section is that HAZY's incremental techniques for maintaining classification views are more efficient than alternate non-incremental techniques. Specifically, we explore the performance of our strategies for: (1) *Updates*, (2) *Single Entity* reads, and (3) *All Members* queries. We perform all combinations of experiments using all five techniques, hazy and naïve main memory (MM), hazy and naïve on-disk (OD), and hybrid for all three operations on three data sets: Forest, DBLife, and Citeseer. We summarize the most interesting experimental results in this section. Finally, in these experiments we restrict the hybrid to hold less than 1% of the entities in main memory (a breakdown of the hybrid's memory usage is in Figure 6(A)).

#### 4.1.1 Improving the Eager Approach

We verify that HAZY's techniques improve the performance of *Update* for an eager approach (which is the slow operation in an eager approach). To assess update performance, we perform the following experiment: we randomly select 3k training examples from each data set. Then, we insert these examples with prepared statements into the database. We then measure the average time per update for each approach. The results are summarized in Figure 4(A).

---
[3]FC is a multiclass classification problem, but we treat FC as a binary classification to find the largest class. We consider the full multiclass classification in Appendix C.

In all cases, the relative standard deviation is less than $10^{-4}$. In our desired applications we have often partially trained a model. To simulate this, the experiment begins with a partially trained (warm) model (after 12k training examples).

The state-of-the-art approach to integrate classification with an RDBMS is captured by the naïve on-disk approach; HAZY's main memory approach achieves between 36x and 124x speedup over this approach. We can also see that using a naïve strategy in memory is an order of magnitude slower than HAZY, which validates our claim that HAZY's strategy is a critical ingredient in our speed up. Interestingly, we see that HAZY is an order of magnitude faster even using an on-disk architecture: HAZY is 8-10x faster on Forest in both main-memory and on-disk architectures. An exception to this observation is Citeseer: we dug deeper, and verified that since Citeseer has a much larger feature space, the model has not converged by the end of the run. Thus, HAZY guesses incorrectly and pays a high cost for sorting. In memory, sorts are very inexpensive, and HAZY achieves a 4x improvement. We also run an experiment that begins with zero examples; here, HAZY still offers orders of magnitude improvement: 111x speedup for Forest, a 60x speedup for DBLife, and a 22x speedup for Citeseer.

#### 4.1.2 Improving the Lazy Approach

In the lazy approach, the bottleneck is the *All Members* query. To assess the extent to which HAZY improves the performance of the *All Members* in a lazy approach, we repeatedly issue a query that asks *"how many entities with label 1 are there?"* We verified that PostgreSQL is using a sequential scan on each iteration (HAZY also scans all entities). To make the comparison between the different approaches fair, we call HAZY via a user-defined function. The state of the art is naïve on-disk, and we see speed ups of over 225x to 525x for all strategies and architectures. Logging reveals that HAZY scans fewer entities and that is the reason that HAZY is faster than the naïve in-memory technique.

*Updates.* *Update* in a lazy approach performs exactly the same code for all architectures and strategies (update the model and return), and so have identical speed: for Forest is 1.6k updates/sec, DBLife is 2.8k updates/sec, and Citeseer is 2.5k updates/sec.

### 4.2 Performance of the Hybrid Architecture

Recall from the cost model, that the main technical claim of the hybrid is that it speeds up *Single Entity* reads. The buffer size of the hybrid architecture is set to 1% of the data size. In this experiment, we ask for the label of 15k uniformly randomly selected entities. The hybrid architecture shines: We see that the hybrid's read rate is 97% of the read

| Arch | Eager | | | Lazy | | |
|---|---|---|---|---|---|---|
| | FC | DB | CS | FC | DB | CS |
| OD | 6.7k | 6.8k | 6.6k | 5.9k | 6.3k | 5.7k |
| Hybrid | 13.4k | 13.0k | 12.7k | 13.4k | 13.6k | 12.2k |
| MM | 13.5k | 13.7k | 12.7k | 13.4k | 13.5k | 12.2k |

**Figure 5:** *Single Entity* **Read. (Read/s)**

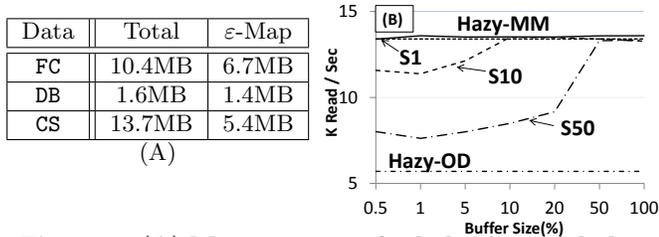

| Data | Total | $\varepsilon$-Map |
|---|---|---|
| FC | 10.4MB | 6.7MB |
| DB | 1.6MB | 1.4MB |
| CS | 13.7MB | 5.4MB |

(A)

**Figure 6: (A)** Memory usage for hybrid on each data set. **(B)** *Single Entity* read v. hybrid buffer size.

rate of the pure main-memory architecture while maintaining only 1% of the data in memory. We list the raw RAM usage of the hybrid in Figure 6. The hybrid is always faster than on-disk architecture for both eager and lazy (hazy and naïve strategies have essentially identical performance); and it has low memory usage. Recall that the $\varepsilon$-map takes the primary key of each entity to its $\varepsilon$ value in the stored model. We see the amount of memory used for the $\varepsilon$-map is much smaller than the entire data set, because it does not contain the feature vector (confirming our assertion in Section 3.5.2).

To understand the effect of the buffer size on overall performance, in Figure 6(B), we vary the buffer size (percentage of entities in the buffer) for three different models that have 1%, 10%, and 50% of the tuples between low and high water (we call them $S1$, $S10$ and $S50$, respectively). Then, we measure the *Single Entity* read rate. We observe that when the buffer size is larger than the number of tuples between low and high water, we achieve close to main-memory architecture performance. By instrumenting the code, we verified that most of the speed up is due to Hazy's ability to quickly prune using the $\varepsilon$-map. Of course, disk access are expensive and we need to have very high hit rate to compensate. When the model has 10% between low and high water, approximately 90% of accesses are answered by this filter. Still, mitigating approximately half the disk access has a noticeable difference: when we keep 5% of the entities in the hybrid's buffer, we get a sizable boost over the on-disk architecture. And so the hybrid's high performance on DBLife is because the average number of tuples between low and high water is small: 4811 of 122k ($\sigma = 122$).

*Extended Experiments.* In Appendix C, we describe experiments on scalability and scale-up (number of threads), and we discuss how we use Hazy in a multiclass problem.

## 5. CONCLUSION

We demonstrate the Hazy approach to incremental maintenance of classification views using the abstraction of model-based views. We show that our algorithmic techniques are able to improve prior art in both the eager and lazy approaches by two orders of magnitude: roughly, one order of magnitude from Hazy's novel algorithm and one from the architectural improvements. The key technical contributions are (1) an incremental algorithm that improves the performance of both eager and lazy approaches inside an RDBMS and (2) a novel hybrid index structure. We present experimental evidence on several datasets.

*Acknowledgements.* This work is supported by the Air Force Research Laboratory (AFRL) under prime contract no. FA8750-09-C-0181, the National Science Foundation under IIS-1054009, and gifts from Google and Johnson Controls, Inc. Any opinions, findings, conclusions, or recommendations expressed in this work do not necessarily reflect the views of DARPA, AFRL, or the US government.

## 6. REFERENCES
[1] L. Antova, T. Jansen, C. Koch, and D. Olteanu. Fast and simple relational processing of uncertain data. In *ICDE*, 2008.
[2] O. Benjelloun, A. D. Sarma, A. Y. Halevy, M. Theobald, and J. Widom. Databases with uncertainty and lineage. *VLDB J.*, 17(2):243–264, 2008.
[3] J. A. Blakeley, P.-Å. Larson, and F. W. Tompa. Efficiently updating materialized views. In *SIGMOD*, pages 61–71, 1986.
[4] L. Bottou. Stochastic graident descent code.http://leon.bottou.org/projects/sgd.
[5] C. J. C. Burges. A tutorial on support vector machines for pattern recognition. *Data Min. Knowl. Discov.*, 2(2), 1998.
[6] G. Cauwenberghs and T. Poggio. Incremental and decremental support vector machine learning. In *NIPS*, pages 409–415, 2000.
[7] S. Chaudhuri, V. R. Narasayya, and S. Sarawagi. Extracting predicates from mining models for efficient query evaluation. *ACM Trans. Database Syst.*, 29(3):508–544, 2004.
[8] C. Cortes and V. Vapnik. Support-vector networks. *Machine Learning*, 20(3):273–297, 1995.
[9] K. Crammer, O. Dekel, J. Keshet, S. Shalev-Shwartz, and Y. Singer. Online passive-aggressive algorithms. *JMLR*, 2006.
[10] N. N. Dalvi and D. Suciu. Efficient query evaluation on probabilistic databases. In *VLDB*, pages 864–875, 2004.
[11] A. Deshpande and S. Madden. MauveDB: supporting model-based user views in database systems. In *SIGMOD*, pages 73–84, 2006.
[12] A. Doan and Others. http://dblife.cs.wisc.edu.
[13] X. Dong, A. Halevy, and J. Madhavan. Reference reconciliation in complex information spaces. In *SIGMOD*, pages 85–96, 2005.
[14] J. Han. *Data Mining: Concepts and Techniques*. Morgan Kaufmann Publishers Inc., San Francisco, CA, USA, 2005.
[15] R. Jampani, F. Xu, M. Wu, L. L. Perez, C. M. Jermaine, and P. J. Haas. MCDB: a monte carlo approach to managing uncertain data. In *SIGMOD*, pages 687–700, 2008.
[16] T. Joachims. Making large-scale SVM learning practical. In *Advances in Kernel Methods - Support Vector Learning*, chapter 11. MIT Press, Cambridge, MA, 1999.
[17] B. Kanagal and A. Deshpande. Online filtering, smoothing and probabilistic modeling of streaming data. In *ICDE*, 2008.
[18] A. R. Karlin, M. S. Manasse, L. A. McGeoch, and S. Owicki. Competitive randomized algorithms for non-uniform problems. In *SODA '90*, Philadelphia, PA, USA, 1990. SIAM.
[19] P. Laskov, C. Gehl, S. Krüger, and K.-R. Müller. Incremental support vector learning: Analysis, implementation and applications. *JMLR*, 7:1909–1936, 2006.
[20] E. Leopold and J. Kindermann. Text categorization with support vector machines. how to represent texts in input space? *Mach. Learn.*, 46(1-3):423–444, 2002.
[21] N. Littlestone and M. K. Warmuth. The weighted majority algorithm. *Inf. Comput.*, 108(2):212–261, 1994.
[22] B. L. Milenova, J. S. Yarmus, and M. M. Campos. Svm in oracle database 10g: removing the barriers to widespread adoption of support vector machines. In *VLDB*, 2005.
[23] C. Ré, J. Letchner, M. Balazinska, and D. Suciu. Event queries on correlated probabilistic streams. In *SIGMOD*, 2008.
[24] W. Rudin. *Principles of mathematical analysis*. McGraw-Hill Book Co., New York, third edition, 1976. International Series in Pure and Applied Mathematics.
[25] S. Shalev-Shwartz, Y. Singer, and N. Srebro. Pegasos: Primal estimated sub-gradient solver for svm. In *ICML*, 2007.
[26] L. Wasserman. *All of Nonparametric Statistics (Springer Texts in Statistics)*. Springer-Verlag New York, Inc., 2006.
[27] F. Wu and D. S. Weld. Automatically refining the wikipedia infobox ontology. In *WWW*, pages 635–644, 2008.

# APPENDIX

## A. MATERIAL FOR SECTION 2

### A.1 Convex Optimization Problem for SVMs

The best model $(\boldsymbol{w}, b)$ is defined via an optimization problem [5]. For $d, p > 0$ and $\boldsymbol{x} \in \mathbb{R}^d$, we define $\|\boldsymbol{x}\|_p^p = \sum_{i=1,\ldots,d} |x_i|^p$. The convex optimization problem is:

$$\min_{\substack{\boldsymbol{w}, b \\ \eta_t: t \in E^+ \cup E^-}} \frac{1}{2} \|w\|_2^2 + C \sum_{t \in E^+ \cup E^-} \zeta_t$$

$$\text{s.t.} \quad \begin{aligned} \boldsymbol{f}(t) \cdot \boldsymbol{w} + b &\leq -1 + \zeta_t & \forall t \in E^- \\ \boldsymbol{f}(t) \cdot \boldsymbol{w} + b &\geq 1 - \zeta_t & \forall t \in E^+ \end{aligned}$$

Here, $\zeta_t \geq 0$ for $t \in T$ are slack variables that allow the classification to be noisy, i.e., we do not assume the data are *linearly separable* [5].

### A.2 Feature Functions

In HAZY, feature functions often have two phases: in the first phase, a function computes some statistics about the corpus, and in the second phase, using that statistical information, a particular tuple is translated into a vector. Thus, we define a feature function as a triple of user-defined functions: `computeStats`, which takes as input a table name and computes whatever information is needed from the whole corpus; `computeStatsInc`, which takes a tuple and incrementally computes statistics; and `computeFeature` that conceptually takes a tuple and the statistical information and returns a vector. In HAZY, these functions communicate via a catalog-like table. An example feature function that requires the catalog is `tf_idf_bag_of_words` computes tf-idf scoring for tuples (treating each tuple as a document) in which the `computeStats` function computes term frequencies and inverse document frequencies for the corpus and stores it in a table, `computeStatsInc` incrementally maintains the inverse document frequencies, and the `computeFeature` function uses this table to compute the tf-idf score for each tuple. In contrast, in `tf_bag_of_words` (term frequencies) no statistics are needed from the corpus. A final example is TF-ICF (term frequency inverse corpus frequency) in which the term frequencies are obtained from a corpus, but explicitly not updated after each new document [31]. In general, an application will know what feature functions are appropriate, and so we design HAZY to be extensible in this regard, but we expect that the administrator (or a similar expert) writes a library of these feature functions. HAZY also supports *kernel functions* that allow support vector machines to handle non-linear classification tasks. Kernels are registered with HAZY as well.

## B. MATERIAL FOR SECTION 3

### B.1 Implementation Details

We implemented HAZY in PostgreSQL 8.4, which has the notable feature of *multi-valued concurrency control* (MVCC). Any attempted in-place write (update) actually performs a copy, and so as an optimization we create a user-defined function that updates records in place without generating a copy. The use of HAZY-MM is transparent to the user: updates are still handled via triggers, and queries against the table are rerouted to HAZY-MM via PostgreSQL's "_RETURN" trigger. To get improved performance, however, HAZY offers several fast-path functions for prepared statements; it is future work to more fully integrate HAZY-MM with PostgreSQL.

**Input**: $a$, the accumulated time, $s$ the sort time.
**Output**: Return $a$
1: **If** $a \geq \alpha S$ **then** REORG(); return 0
2: **Else return** $a$ + TIME(Take_Incremental_Step())

**Figure 7: The Skiing strategy.**

### B.2 Proof of Lemma 3.1

The proof of Lemma 3.1 uses two applications of Hölder's inequality [24, p.139], which says that $|\langle x, y \rangle| \leq \|x\|_p \|y\|_q$ where $p^{-1} + q^{-1} = 1$ and $p, q \geq 1$.

PROOF OF LEMMA 3.1. Denote $t$.eps by $\varepsilon_t$. We write $\boldsymbol{w}^{(i+1)} = \boldsymbol{w}^{(s)} + \boldsymbol{\delta}_w$, $b^{(i+1)} = b^{(s)} + \delta_b$. Fix parameters $p, q \geq 0$ such that $p^{-1} + q^{-1} = 1$ and let $M = \max_t \|f(t)\|_q$. Observe that

$$\begin{aligned} t \in V_+^{(i+1)} &\iff \langle \boldsymbol{w}^{(s)} + \boldsymbol{\delta}_w, t, f \rangle - (b + \delta_b) > 0 \\ &\iff \varepsilon_t > -(\langle \boldsymbol{\delta}_w, f(t) \rangle - \delta_b) \end{aligned}$$

We use Hölder's inequality: $|\langle \delta_w, f(t) \rangle| \leq \|\delta_w\|_p \|f(t)\|_q \leq \|\delta_w\|_p M$. So if $\varepsilon_t \geq M \|\delta_w\|_p + \delta_b$ then $\varepsilon_t \geq -(\langle \boldsymbol{\delta}_w, f(t) \rangle - \delta_b)$ and so $t \in V_+^{(i)}$. On the other hand, if $\varepsilon_t \leq -M \|\delta_w\|_p + \delta_b$ then $t \in V_-^{(i)}$. □

### B.3 Skiing Strategy Analysis

Fix a $\sigma \geq 0$ and let $\alpha$ be the positive root of $x^2 + \sigma x - 1$. The SKIING strategy is: Initially set $a$, the accumulated cost, to 0. At time $i$, if $a \geq \alpha S$ then choose option (2) and set $a$ to 0. Otherwise, select option (1) and set $a$ to $a + c^{(i)}$. Let $\text{OPT}_{\bar{c}}$ denote the optimal solution for $\bar{c}$ over all schedules.

LEMMA B.1. *For any set of costs $\bar{c}$ of the algorithm:*

$$\text{cost}(\text{SKIING}) \leq (1 + \sigma + \alpha)\text{cost}(\text{OPT}_{\bar{c}})$$

PROOF. Consider a run of length $N$ and without loss of generality assume that it ends after an even number of SKIING reorganizations. Suppose that SKIING chooses option 1 at time $0, t_1, \ldots, t_{2M}$. Consider any interval $[t_m, t_{m+2}]$. Let $C_1$ the cost of SKIING on $[t_m, t_{m+1})$ and $C_2 = (t_{m+1}, t_{m+2})$ so that the cost of SKIING is $C_1 + C_2 + 2S$ on this interval. Observe that $\alpha S \leq C_i < (\sigma + \alpha)S$ for $i = 1, 2$. Suppose that OPT does not reorganize in this interval, then it pays at least cost $C_1 + C_2$ and so SKIING's competitive ratio is $\frac{1+\alpha}{\alpha}$ on this interval. Suppose that OPT reorganizes once, then its cost is at least $\min\{C_1, C_2\} + S$. Without loss let $zS = C_1$ then our claim is that: $\frac{2+z+\alpha+\sigma}{1+z} \leq 1 + \alpha + \sigma$, which reduces to the claim: $0 \leq z(\sigma + \alpha) - 1$. Now, $z \geq \alpha$, when $z = \alpha$ then the right hand side is 0 by the definition of $\alpha$; then observe that the rhs is an increasing function in $z$, which completes the claim. Finally, suppose that OPT reorganizes $k \geq 2$ times, then it pays at least $kS$, and since $C_1 + C_2 \leq 2(\alpha + \sigma)S$, in this case the ratio is $\frac{2(1+\alpha+\sigma)S}{kS}$. Since $k \geq 2$, the largest this ratio can be is $1 + \alpha + \sigma$. □

We prove the following lower bound.

THEOREM B.2. *For any deterministic strategy $\Psi$ where $\sigma$ and $\alpha$ set as above*

$$\rho(\Psi) \geq 1 + \sigma + \alpha$$

PROOF. We assume without loss that the approximation factor is not infinite, otherwise the theorem trivially holds, i.e. there is some $B \geq 0$ such that $\mathrm{Cost}(\Psi) \geq B \cdot \mathrm{Cost}(\mathrm{OPT}_{\bar{c}})$ for all $\bar{c}$. Fix $\varepsilon > 0$. Consider the following family of costs: $c^{(k,i)}(z) = \varepsilon$ if $z = 0$ and $k \leq i$. For some $i$, $\Psi$ must reorganize call $t$ is this index – otherwise its approximation is unbounded, the optimal simply reorganizes first and then incurs cost 0.

We first observe that if the reorganization occurs at $t$ then it has paid cost $t\varepsilon = \beta S$ for some $\beta$ and so $\Psi$ pays $(1+\beta)S$. The optimal solution here does not switch at $t$ and pays cost $\beta S$ so the ratio is at least $\frac{1+\beta}{\beta}$. (This is loose if $\beta \geq 1$, but it does not matter for the theorem).

Now create a second set of costs such that: $\bar{d} = \bar{c}$ except that $d^{(t,t)}(0) = S$. Since $\Psi$ is a deterministic online strategy it produces the same sequence on this set of costs. If $\Psi$ does not reorganize at $t$, then the adversary can continue to make $\Psi$ pay non-zero cost, while the optimal pays 0 cost, until $\Psi$ chooses to reorganize. As a result, $\Psi$ pays at least $(1+\beta+\sigma)S - \varepsilon$. In contrast, the optimal reorganizes once and pays $S$, and so the ratio is $(1+\sigma+\beta) - \varepsilon S^{-1}$. Letting $\varepsilon \to 0$, the competitive ratio satisfies:

$$\max_{\bar{c}} \frac{\mathrm{Cost}(\Psi; \bar{c})}{\mathrm{Cost}(\mathrm{OPT}; \bar{c})} \geq \max\left\{\frac{1+\beta}{\beta}, 1+\sigma+\beta\right\}$$

Now, using the fact that for $\beta \geq 0$ we have:

$$\max\left\{\frac{1+\beta}{\beta}, 1+\sigma+\beta\right\} \geq 1 + \frac{1}{2}\left(\sigma + \sqrt{\sigma^2 + 4}\right)$$

and this minimum is achieved at $\beta$ is the positive root of $x^2 + \sigma x - 1$, which is when $\alpha = \beta$. □

*Non-monotone Incremental Step Functions.* We can optimize our choice of incremental function, which makes our low and high water violate our property (2) from Section 3.3.

$$lw^{(s,i)} = \min_{j=i-1,i} \varepsilon_{low}^{(s,j)} \text{ and } hw^{(s,i)} = \max_{j=i-1,i} \varepsilon_{high}^{(s,j)}$$

Experimentally, we have validated that the cost differences between the two incremental steps is small. Still, theoretically, it is natural to wonder if our analysis can be applied to the case without the monotonicity assumption. Informally, if we drop the monotonicity assumption, then no approximation factor is possible. Consider any deterministic strategy on an input that no matter where its incremental cost is 1. Suppose the strategy never reorganizes, then consider an optimal strategy: at stage 1, it reorganizes and then pays cost 0 from then on. Its total cost is 1, but the strategy that never reorganizes has unbounded cost. On the other hand, suppose that the strategy reorganizes at least once on long enough input. Consider an optimal strategy that reorganizes at any other time, then, the optimal pays a fixed cost $S$ while the strategy is again pays an unbounded cost.

**Input**: id (a tuple id), $h^{(s)}$ hybrid's $\varepsilon$-map, $lw^{(s,i)}$, and $hw^{(s,i)}$.
**Output**: Return $+1$ if $t$ is in the class and $-1$ otherwise.
1: **If** $h^{(s)}(t) \leq lw_{low}^{(s,i)}$ **then** return $-1$
2: **If** $h^{(s)}(t) \geq hw_{low}^{(s,i)}$ **then** return $+1$
3: **If** id in buffer **then** return class
4: **else** look up class using on-disk HAZY.

Figure 8: Hybrid Look up Algorithm

| Name | $L(z,y)$ | Name | $P(\boldsymbol{w})$ |
|------|----------|------|---------------------|
| SVM | $\max\{1-zy, 0\}$ | $\ell_p$ ($p \geq 1$) | $\lambda \|\boldsymbol{w}\|_p^p$ |
| Ridge | $(z-y)^2$ | Tikhanov | $\lambda \|\boldsymbol{w}\|_Q$ |
| Logistic | $\log(1+\exp(-yz))$ | KL | $-H(\boldsymbol{w})$ |
| (a) | | (b) | |

Figure 9: (a) Common loss functions. (b) Common penalty (regularization) terms. Here we abbreviate $z = \boldsymbol{x} \cdot \boldsymbol{w}$, $\|\boldsymbol{w}\|_Q = \boldsymbol{w}^T Q \boldsymbol{w}$ for a positive definite $Q$, and the binary entropy as $H(\boldsymbol{w})$.

### B.4 Hybrid Approach

Figure 8 summarizes the search routine for the hybrid data structure.

### B.5 Extensions to the Model

We discuss here how many extensions are easy to integrate into our model in HAZY; we believe this illustrates a strength of our architecture.

#### B.5.1 Other Linear Classification Methods

So far for concreteness, we have focused on *support vector machines*, which are one of the most popular forms of linear models. Other linear models differ from SVMs in that they optimize different metrics. Most popular classification models can be written as convex (but non-smooth) optimization problems in the following form:

$$\min_{\boldsymbol{w}} P(\boldsymbol{w}) + \sum_{(x,y) \in T} L(\boldsymbol{w} \cdot \boldsymbol{x}, y)$$

where $P$ is a strongly convex function called a *regularizer* and $L$ is a convex function called the *loss function*. Figure 9 lists $P$ and $L$ for some popular linear models that are supported by HAZY. Internally, HAZY uses (incremental) gradient-based methods, and we have found it very easy to add new linear models. Typically, a new linear requires tens of lines of code.

The reason that these models are called linear is that the label of a function $\boldsymbol{x}$ (denoted $l(\boldsymbol{x})$) depends on the value of $\boldsymbol{w} \cdot \boldsymbol{x}$ where $\boldsymbol{w}$ is a model:

$$l(\boldsymbol{x}) = h(\boldsymbol{w} \cdot \boldsymbol{x})$$

For example, in SVMs, ridge regression, logistic regression, $h$ is simply the sign function. The only property we use in our previous algorithm is that $h$ is monotone non-decreasing.

#### B.5.2 Kernel Methods

Kernel methods extend linear methods using the *kernel trick* [5]. A kernel $K: \mathbb{R}^d \times \mathbb{R}^d \to \mathbb{R}$ is a positive semi-definite function.[4] We can write a kernel classifier as:

---
[4]A function $K: \mathbb{R}^d \times \mathbb{R}^d \to \mathbb{R}$ is positive semi-definite

311

$$c(\boldsymbol{x}) = \sum_{i=1,\ldots,N} c_i \cdot K(\boldsymbol{s}_i, \boldsymbol{x})$$

here each $c_i$ is a real-valued weights, and $\boldsymbol{s}_i$ is called a *support vector*. The number of support vectors, $N$, may be on the order of the number of training examples, but can be substantially smaller.

Here, the model is encapsulated by the weights on the support vectors. The same intuition still holds: if $w + \delta = w'$, then observe that all the above kernels $K(s_i, \boldsymbol{x}) \in [0, 1]$ hence the maximum difference is the $\ell_1$ norm of $\delta$. Then, we can apply exactly the same algorithm. Here, we have regarded both models as being in the same space (the first model assigns 0 weight to the new training example). This is important as a new training example can introduce a new support vector.

### B.5.3 Linearized Kernels

For shift-invariant kernels[5], such as the Gaussian and the Laplacian kernel, we directly reuse our linear techniques from the body of the paper by transforming these kernels to (low) dimensional linear spaces.

The idea is based on a technique of Rahimi and Recht called *random non-linear feature vectors* [30]. Suppose that all vectors are in $\mathbb{S}^d$, the unit ball in $d$ dimensions (any compact set will do). The idea is to find a (random) map $z : \mathbb{S}^d \to \mathbb{R}^D$ for some $D$ such that for $x, y \in \mathbb{S}^d$ we have $z(x)^T z(y) \approx K(x, y)$. More precisely, with arbitrarily high probability, we find a function $z$ such that $|z(x)^T z(y) - K(x, y)| \leq \varepsilon$ for *all* $x, y \in \mathbb{S}^d$ *simultaneously*. The value of $D$ depends directly on the confidence we want in this property and $\varepsilon^{-1}$. We can describe one such $z$ explicitly: Choose $\boldsymbol{r}^{(i)} \in \mathbb{R}^{d+1}$ uniformly distributed on the sphere, then the $i$th component $z(\boldsymbol{x})$ is given by $z(\boldsymbol{x})_i = \sqrt{2}\cos(\boldsymbol{r}^{(i)} \cdot \boldsymbol{x})$.

Given such a $z$, observe that the kernel function $f(x) = \sum_{i=1,N} \alpha_i K(s_i, x) \approx \sum_{i=1,N} \alpha_i z(s_i)^T z(x_i) = v^T z(x)$ where $v = \sum_{i=1,N} \alpha_i z(s_i)$. This can be a substantial reduction: $|v| = D$ and so testing a point is $O(D)$ versus $O(Nd)$. Also, given $z$ we can transform the learning and testing problem to a linear problem. Often, this problem is indistinguishable in quality from the original kernel. Importantly for this work, the problem is again a linear classification problem.

### B.5.4 Multiclass Classification

One standard technique to turn a binary classification into a multiclass classification is to build a decision-tree-like structure where each node corresponds to a single binary SVM. HAZY supports multiclass classification with similar efficiency gains for this case.

## C. EXTENDED EXPERIMENTS

## C.1 Overhead of Learning and Classification

In HAZY, the RDBMS performs all data manipulation for learning and classification. Of course, this can have overhead compared to hand-tuned solutions using files. To quantify this overhead, we compare HAZY's solution where each

---

if for any $x_1, \ldots, x_N \in \mathbb{R}^d$, $c_1, \ldots, c_N \in \mathbb{R}$, we have $\sum_{i,j=1,\ldots,N} K(x_i, x_j) c_i c_j \geq 0$.

[5]A function $f(x, y)$ is *shift invariant* if it can be written $f(x - y)$. A kernel is shift invariant if it is shift invariant regarded as a function.

| Data set | SVMLIGHT | | SGD-based | | |
|---|---|---|---|---|---|
| | P/R | Time | P/R | File | HAZY |
| MAGIC | 74.4/63.4 | 9.4s | 74.1/62.3 | 0.3s | 0.7s |
| ADULT | 86.7/92.7 | 11.4s | 85.9/92.9 | 0.7s | 1.1s |
| FOREST | 75.1/77.0 | 256.7m | 71.3/80.0 | 52.9s | 17.3m |

**figure 10:** Performance of SVMLight, a stochastic gradient method (no RDBMS) [4], and Hazy.

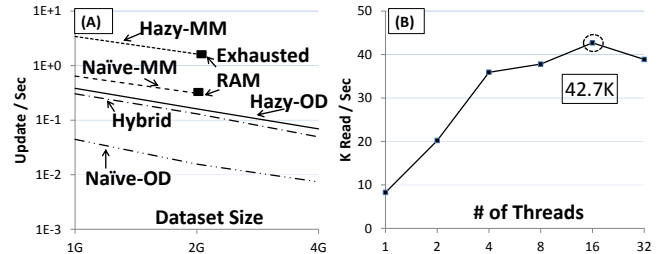

**Figure 11:** (A) Scalability. (B) Scaleup.

training example is a single update statement against two popular approaches in Figure 10: (1) a stochastic gradient method (SGD) [4] where all data manipulations are done in hand coded C-files, and (2) SVMLIGHT [16], a popular package to do SVM classification that has been tuned over several years. In Figure 10, we compare our approach on several standard machine learning data sets with 90% as training data. Overall, we see that the SGD-based approach is much faster than SVMLIGHT and has essentially the same quality. The overhead that HAZY imposes over the hand-tuned SGD-based solution is due primarily to the overhead of PostgreSQL to do insert-at-a-time update method. To verify this, we implemented a bulkloading method, and the time dropped to 44.63s to classify Forest (parsing code in HAZY is slightly optimized).

## C.2 Scalability, Scale-up, and Sensitivity

Figure 11(A) shows a data scalability experiment for eager updates. We created a synthetic data sets of sizes 1GB, 2GB and 4GB ($x$-axis). We then plot the eager update performance with a warm model: as we can see, the HAZY main memory technique has the best update performance, but is unable to run on the 4GB data set (since the test machine has 4GB of RAM). We see that HAZY on-disk is close to the naïve main memory technique and scales well with the data set size. Additionally, we see that the hybrid architecture pays only a small penalty on update over the on disk architecture (it has a more expensive resort). We conclude that HAZY scales well in the data set size.

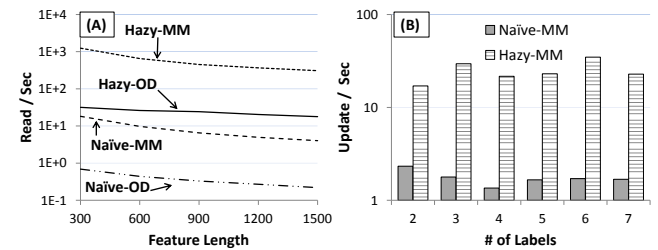

**Figure 12:** (A) Feature Sensitivity. (B) Multiclass.

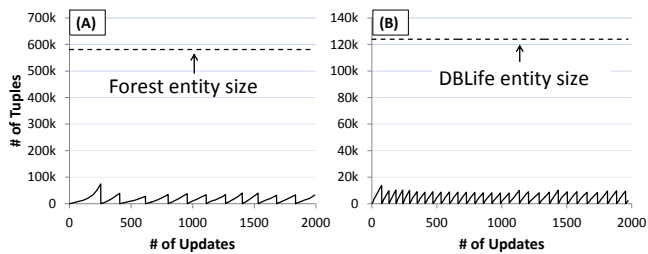

**Figure 13: Tuples between Low Water and High Water on (A) Forest and (B) DBLife.**

One feature of our approach is that the locking protocols are trivial for *Single Entity* reads. To demonstrate this fact, we plot our read performance for *Single Entity* in main memory as we scale up the number of threads. We see that slightly over-provisioning the threads to compensate for PostgreSQL latency achieves the best results; this experiment is run on an 8 core machine and 16 threads achieves peak performance (42.7k reads/sec). For comparison, training forest on this machine takes 195.22 minutes.

We conduct two sensitivity experiments: (1) $\alpha$-sensitivity (the parameter in SKIING) and (2) feature sensitivity graphs by scaling up the size of the features. We found that by tuning $\alpha$ we could improve the performance approximately 10% over the default choice setting of $\alpha = 1$. In Figure 12, we conduct an experiment for (2): We scaled up the number of features (using the random features of Appendix B.5.3) and measured the performance of the *All Members* query in lazy HAZY and lazy naïve for main-memory and on-disk architectures. We found that HAZY excels in these situations as it avoids dot-products which have become more costly. We conclude that HAZY scales well with the data set size, is able to take advantage of multicore machines, and is insensitive to feature size.

One key piece of intuition in the HAZY algorithm is that the number of tuples between low water and high water will be a small fraction of the overall tuples. We confirm this intuition in Figure 13(A) and Figure 13(B) with the following experiment: we count the number of tuples that are between low and high water after we have seen 12k update examples (warm model). Indeed, in the steady state we see that approximately 1% of the tuples are between low and high water on both Forest and DBLife.

## C.3 Extension: Multiclass Classification

We train a multiclass configuration using several different configurations of binary classification. We present only a sequential one-versus-all approach for performance. In Figure 12, we vary the number of classes and measure eager update performance in the classification to understand how HAZY is affected. The model is warm with 12k training examples. The data set is `Forest` where we have manually coalesced some of classes together. We see that HAZY maintains its order of magnitude advantage in updates per second over a naïve memory technique as the number of classes increases, which suggests that HAZY offers value in multiclass settings.

## D. EXTENDED RELATED WORK

Incorporating support vector machines into database management systems is an ongoing trend: Oracle 11g supports support vector machines, but they do not consider incremental training nor do they consider incremental maintenance of results.

Data intensive and semantically rich problems, such as querying data from information extraction or data integration, often produce data that can be contradictory and imprecise; this observation has motivated research to treat this data as *probabilistic databases* [1, 2, 10] and several sophisticated statistical frameworks have been developed such as *Factor Graphs* [33], and *the Monte Carlo Database* [15].

An interesting problem that is related, but orthogonal, problem is *active learning* [32, 35], where the goal is to leverage the user feedback to interactively build a model. Technically, our goal is to solicit feedback (which can dramatically help improve the model). In fact one of our initial motivations behind the hybrid approach is to allow active learning over large data sets.

There has been research on how to optimize the queries on top of data mining predicates [28]; we solve the complementary problem of incrementally maintaining the output of the models as the underlying data change. There has been work on maintaining data mining models incrementally, notably association rule mining [29, 34, 36, 37], but not for the general class of linear classifiers. Researchers have considered scaling machine learning tool kits that contain many more algorithms than we discuss here, e.g., WekaDB [38]. These approaches may result in lower performance than state-of-the-art approaches. In contrast, our goal is to take advantage of incremental facilities to achieve higher levels of performance.

## E. REFERENCES

[28] S. Chaudhuri, V. Narasayya, and S. Sarawagi. Efficient evaluation of queries with mining predicates. In *ICDE*, pages 529–540, 2002.
[29] D. Cheung, J. Han, V. Ng, and C. Wong. Maintenance of discovered association rules in large databases: An incremental updating technique. In *ICDE*, 1996.
[30] A. Rahimi and B. Recht. Random features for large-scale kernel machines. In *NIPS*, 2007.
[31] J. Reed, Y. Jiao, T. Potok, B. Klump, M. Elmore, and A. Hurson. Tf-icf: A new term weighting scheme for clustering dynamic data streams. In *ICMLA*, 2006.
[32] S. Sarawagi and A. Bhamidipaty. Interactive deduplication using active learning. In *KDD*, pages 269–278, 2002.
[33] P. Sen, A. Deshpande, and L. Getoor. Prdb: managing and exploiting rich correlations in probabilistic databases. *VLDB J.*, 18(5):1065–1090, 2009.
[34] J. C. Shafer, R. Agrawal, and M. Mehta. Sprint: A scalable parallel classifier for data mining. In *VLDB*, 1996.
[35] S. Tong and D. Koller. Support vector machine active learning with applications to text classification. *JMLR*, pages 45–66, 2001.
[36] A. Veloso, W. M. Jr., M. de Carvalho, B. Pôssas, S. Parthasarathy, and M. J. Zaki. Mining frequent itemsets in evolving databases. In *SDM*, 2002.
[37] A. Veloso, B. G. Rocha, W. M. Jr., M. de Carvalho, S. Parthasarathy, and M. J. Zaki. Efficiently mining approximate models of associations in evolving databases. In *PKDD*, pages 435–448, 2002.
[38] B. Zou, X. Ma, B. Kemme, G. Newton, and D. Precup. Data mining using relational database management systems. In *PAKDD*, pages 657–667, 2006.